\documentclass[prl,twocolumn,showpacs,preprintnumbers,amsmath,amssymb]{revtex4}
\usepackage{graphicx}
\begin{document}
\title{
Phonon-induced Exciton Dephasing in Quantum Dot Molecules}
\author{
E. A. Muljarov$^{1,2,3}$}\email{muljarov@gpi.ru}
\author{T. Takagahara$^2$ }
\author{R. Zimmermann$^1$}
\affiliation{ $^1$Institut f\"ur Physik der Humboldt-Universit\"at
zu Berlin,
Newtonstrasse 15, D-12489 Berlin, Germany\\
$^2$Kyoto Institute of Technology, Matsugasaki, Sakyo-ku, Kyoto 606-8585, Japan  \\
$^3$General Physics Institute, Russian Academy of Sciences,
Vavilova 38, Moscow 119991, Russia
 }
\begin{abstract}
A new microscopic approach to the optical transitions in quantum
dots and quantum dot molecules, which accounts for both diagonal
and non-diagonal exciton-phonon interaction, is developed. The
cumulant expansion of the linear polarization is generalized to a
multilevel system and is applied to calculation of the full time
dependence of the polarization and the absorption spectrum. In
particular, the broadening of zero-phonon lines is evaluated
directly. It is found that in some range of the dot distance real
phonon-assisted transitions between exciton states dominate the
dephasing,  while virtual transitions are of minor importance. The
influence of Coulomb interaction, tunneling, and structural
asymmetry on the exciton dephasing in quantum dot molecules is
analyzed.
\end{abstract}
\date\today

\maketitle

Semiconductor quantum dots (QDs) have been considered for several
years as promising candidates to form an elementary building block
(qubit) for quantum computing. Recently quantum dot molecules
(QDMs), i.e. systems of two quantum-mechanically coupled QDs, have
been proposed for realization of optically driven quantum gates
involving two-qubit operations~\cite{Burkard99,Bayer01,Ortner05}.
It turns out, however, that contrary to their atomic counterparts,
QDs have a strong temperature-dependent dephasing of the optical
polarization. Such a decoherence caused by the interaction of the
electrons with the lattice vibrations (phonons) is inevitable in
solid state structures thus presenting a fundamental obstacle for
their application in quantum computing.

There has been considerable progress in the understanding of
exciton dephasing in QDs after a seminal publication by Borri~{\it
et al.} on four-wave mixing measurements in InGaAs
QDs~\cite{Borri01}. Two important and well understood features of
the dephasing are: (i) the optical polarization experiences a
quick initial decay within the first few picoseconds after pulsed
excitation and (ii) at later times it shows a much slower
exponential decay. In photoluminescence and absorption spectra
this manifests itself as (i) a broadband and (ii) a much narrower
Lorentzian zero-phonon line (ZPL) with a temperature dependent
linewidth~\cite{Besombes01}. Such a behavior of the polarization
is partly described within the widely used independent boson
model~\cite{Mahan} that allows an analytic solution for the case
of a single exciton state. It describes satisfactorily the
broadband (or the initial decay of the polarization). However, in
this model there is no long-time decay of the polarization (no
broadening of the ZPL).

Recently we have presented a first microscopic calculation of the
ZPL width in single QDs~\cite{Muljarov04}, taking into account
{\it virtual} phonon-assisted transitions into higher exciton
states and mapping the off-diagonal linear exciton-phonon coupling
to a diagonal but quadratic Hamiltonian. This is the major
mechanism of phonon-induced dephasing in single QDs as long as the
exciton level distance is much larger than the typical energy of
the acoustic phonons coupled to the QDs ($\leq3$\,meV in InGaAs
QDs). On the contrary, in QDMs the distance between the nearest
exciton levels can be made arbitrary small if the tunnelling
between dots is weak enough~\cite{Szafran01}, so that the
interaction with acoustic phonons can lead to {\it real}
transitions (changing the level occupation). The experimentally
measured excitonic polarization shows a different behavior in
QDMs~\cite{Borri03}, too: the long-time decay is
multi-exponential, in contrast to a single-exponential one in
uncoupled QDs~\cite{Borri01}.

To take into account both real and virtual transitions we develop
in this Letter a new approach for a multilevel excitonic system
which is coupled to acoustic phonons both diagonally and
non-diagonally. This allows us to calculate the dephasing in QDMs,
as well as the full time-dependent linear polarization and
absorption. Instead of the self-energy
approach~\cite{Takagahara99} which is more standard in the
electron-phonon problem, we use the cumulant
expansion~\cite{Kubo62,Mahan}. It is much more advantageous when
studying the dephasing and, being applied to a multilevel system,
has to be generalized to a matrix form.

To calculate the linear polarization we reduce the full excitonic
basis to the Hilbert space of single exciton states $|n\rangle$
(with bare transition energies $E_n$ and wave functions $\Psi_n$).
Then the exciton-phonon Hamiltonian takes the form
\begin{eqnarray}
&&H=\sum_n E_n |n\rangle \langle n|+\sum_{\bf q} \omega_{q}
a^\dagger_{\bf q} a_{\bf q}+\sum_{nm}V_{nm} |n\rangle \langle m|
\nonumber\\
&&V_{nm}=\sum_{\bf q} M^{nm}_{\bf q} (a_{\bf q}+a^\dagger_{-\bf
q}),\label{H}
\\
&&M^{nm}_{\bf q}=\sqrt{\frac{\omega_q}{2\rho_M u_s^2 {\cal
V}}}\int d{\bf r}_e d{\bf r}_h  \Psi^\ast_n({\bf r}_e,{\bf r}_h)
\Psi_m({\bf r}_e,{\bf r}_h)
\nonumber\\
&&\phantom{M^{nm}_{\bf q}=} \times\left[ D_c e^{i{\bf q\,r}_e}-D_v
e^{i{\bf q\,r}_h}\right],\label{M}
\end{eqnarray}
where $a^\dagger_{\bf q}$ is the acoustic phonon creation
operator, $D_{c(v)}$ is the deformation potential constant of the
conduction (valence) band, $\rho_M$ is the mass density, $u_s$ the
sound velocity, ${\cal V}$ the phonon normalization volume, and
$\hbar=1$.

The linear polarization is given by
\begin{equation}
P(t)=\langle \hat{d}(t) \hat{d}(0)\rangle =\sum_{nm} d^\ast_n d_m
 e^{-iE_n t}P_{nm}(t)
 \label{P}
\end{equation}
where $d_n=p_{cv}\int d{\bf r} \Psi_n({\bf r},{\bf r})$ are the
projections of the excitonic dipole moment operator $\hat{d}$
($|d_n|^2$ are the oscillator strengths). The components of the
polarization are written as standard perturbation series,
\begin{eqnarray}
P_{nm}(t)&=&\sum_{k=0}^\infty (-1)^k\int_0^t
dt_1\int_0^{t_1}dt_2\dots \int_0^{t_{2k\!-\!1}} dt_{2k} \nonumber
\\
&&\times \sum_{p\,r\dots\, s}
e^{i\Omega_{np}t_1}e^{i\Omega_{pr}t_2}\dots e^{i\Omega_{sm}t_{2k}}
\nonumber\\
&&\times\langle V_{np} (t_1) V_{pr} (t_2)\dots
V_{sm}(t_{2k})\rangle,  \label{Pnm}
\end{eqnarray}
where the finite-temperature expectation value is taken over the
phonon system, and the difference energies are
$\Omega_{nm}=E_n-E_m$. The expansion of $\exp(-iE_n t)P_{nm}(t)$
which is in fact the full exciton Green's function is shown
diagrammatically in Fig.\,1 up to second order, where the phonon
Green's function $\langle {\cal T}V_{nk} (t) V_{pm} (t')\rangle$
(dashed lines) depends on four exciton indices.

\begin{figure}[b]
\includegraphics*[width=8.0cm]{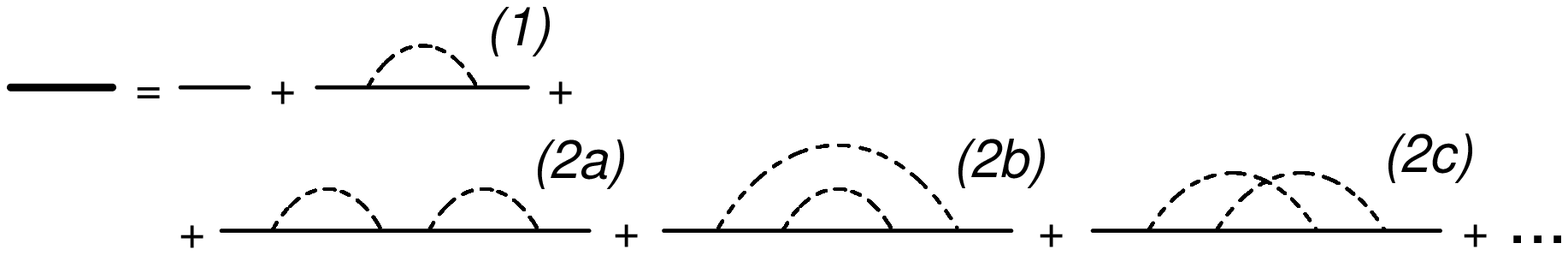}
\caption{ Diagram representation of the perturbation series for
the full exciton Green's function up to second order. }
\end{figure}

Instead of the plain or self-energy summation of diagrams we
introduce for each time $t$ the cumulant matrix $\hat{K}(t)$
defined as
\begin{equation}
\hat{P}(t)=e^{\hat{K}(t)},\label{PexpK}
\end{equation}
where $\hat{P}(t)=\hat{1}+\hat{P}^{(1)}+\hat{P}^{(2)}+\dots$ is
the expansion of the polarization matrix $P_{nm}$ given by
Eq.(\ref{Pnm}). Then the corresponding expansion for the cumulant
is easily generated:
\begin{equation}
\hat{K}(t)=\hat{P}^{(1)}+\hat{P}^{(2)}-\frac{1}{2}
\hat{P}^{(1)}\hat{P}^{(1)}+\dots\label{K}
\end{equation}

Numerically, we restrict ourselves to a finite number of exciton
levels and diagonalize the cumulant matrix $\hat{K}$ at a given
time in order to find the polarization via Eq.(\ref{PexpK}).

If all off-diagonal elements of the exciton-phonon interaction are
neglected ($M_{\bf q}^{nm}=\delta_{nm}M_{\bf q}^{nn}$), the
cumulant expansion ends already in first order: the contribution
of all higher terms of the polarization is exactly cancelled in
the cumulant, Eq.(\ref{K}), by lower order products. This result
allows the exact solution of the independent boson
model~\cite{Mahan}. The inclusion of the non-diagonal interaction
leads to non-vanishing terms in the cumulant in any order. Still,
there is a partial cancellation of diagrams which provides a large
time asymptotics of the cumulant, $\hat{K}(t)\to
-\hat{S}-i\hat{\omega} t -\hat{\Gamma}t$, that is linear in time.
Consequently, the lineshape of the ZPL is Lorentzian. For example,
diagrams {\it (1)}, {\it (2b)}, and {\it (2c)} in Fig.\,1 behave
linear in time at $t\gg L/u_s$ ($L$ is the QD size), while diagram
{\it (2a)} has a leading $t^2$ behavior. In the cumulant, however,
this quadratic term is cancelled exactly by the square of diagram
{\it (1)}, $\hat{P}^{(1)}\hat{P}^{(1)}/2$.

The broadening of the ZPL (which is absent in the independent
boson model) is exclusively due to the non-diagonal exciton-phonon
interaction and appears already in first order of the cumulant.
Remarkably, the cumulant expansion reproduces in first order
exactly Fermi's Golden rule for the {\it real} phonon-assisted
transitions: $\Gamma^{(1)}_1=\pi N_{\rm Bose}(\Delta E)\sum_{\bf
q} |M^{12}_{\bf q}|^2 \delta(\Delta E  -\omega_q)$, where the
ground state dephasing rate $\Gamma_1^{(1)}$ is given here for a
system with two excitonic levels only. It is simulated for a
single spherical QD as a function of the level distance $\Delta
E=E_2-E_1$ (Fig.\,2, dashed curve). As $M_{\bf q}^{12}$ decays
with ${\bf q}$ due to the localization of the exciton wave
functions (Gauss type in the present model calculation),
$\Gamma_1^{(1)}$ also decays quickly with $\Delta E$ (see the
parabola in the logarithmic scale). It exhibits a maximum at
$\Delta E= \omega_0\sim u_s/L$ which is a typical energy of
phonons coupled to the QD.

\begin{figure}[b]
\includegraphics*[width=7.5cm]{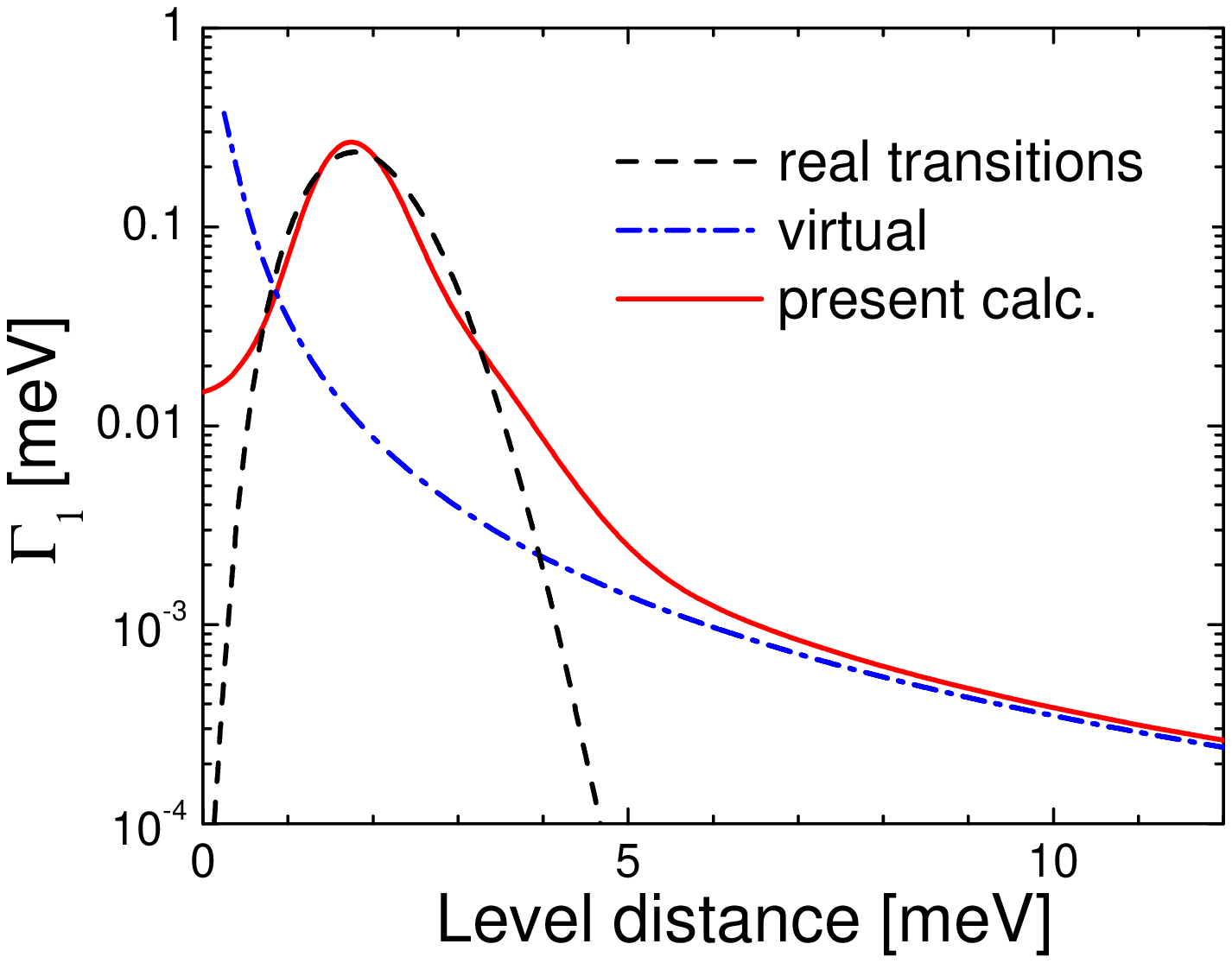}
\caption{ Broadening of the ground state ZPL as a function of the
exciton level distance $E_2-E_1$ in InAs spherical QD, calculated
with account for real transitions (in first order), virtual
transitions (according to Ref.\,\cite{Muljarov04}) and for both
real and virtual transitions up to second order in the cumulant.}
\end{figure}

In spite of this `phonon bottleneck' effect, the  {\it virtual}
transitions are always present in QDs due to second-order diagrams
{\it (2b)} and {\it (2c)} and lead to a non-vanishing broadening
of the ZPL everywhere. They have been taken into account already
within our quadratic coupling model~\cite{Muljarov04} which is
valid in the opposite limit $\Delta E \gg \omega_0$ (dash-dotted
curve in Fig.\,2). In the present calculation we do much better
than in Ref.\,\cite{Muljarov04}: We account for both real and
virtual transitions on an equal footing (up to second order in the
cumulant) and cover the full range of possible values of exciton
level distances (Fig.\,2, full curve).

To describe excitonic states in a QDM, we restrict ourselves in
the present calculation to a four-level model. We take into
account for electron and hole the two lowest localized states
each. Without Coulomb interaction the electron-hole pair state is
a direct product of the one-particle states which form our basis
of four states. Then we include the Coulomb interaction and
diagonalize a four-by-four Hamiltonian. Such a four-level model is
valid as far as the Coulomb matrix elements are smaller than the
energetic distances to higher confined ($p$-shell) or wetting
layer states.

Since a strictly symmetric QDM would lead by degeneracy to special
features~\cite{Szafran01}, we concentrate here on a slightly
asymmetric situation: the confining potentials of the left dot are
2\% deeper than those of the right one. This is close to the
realistic situation, when the In concentration fluctuates from dot
to dot. At the same time, the shape fluctuations of the QDM are
less important for shallow QDs studied in the
experiments~\cite{Bayer01,Ortner05,Borri03}. Thus we assume both
dots to have the same cylindrical form with height $L_z=1$\,nm (in
the growth direction) adjusted from the comparison with
experimentally measured transition energies~\cite{Bayer01,Borri03}
(taking 92\% of In concentration). Given that $s$-, $p$-, $d$-,
and $f$-shells in the luminescence spectra of QDMs have nearly
equidistant positions~\cite{Fafard00}, the in-plain confining
potentials are taken parabolic with Gaussian localization lengths
of carriers adjusted to $l_e=6.0$\,nm and
$l_h=6.5$\,nm~\cite{note1}. The electronic band parameters are
taken from Ref.\,\cite{Landolt} and the acoustic phonon parameters
are the same as used previously~\cite{Muljarov04}.

While in single dots the Coulomb interaction results in a small
correction to the polarization decay, in QDMs the exciton wave
functions (and consequently $M_{\bf q}^{nm}$) are strongly
affected by the Coulomb energy~\cite{Szafran01}. Even more
important is the influence of the Coulomb interaction and
asymmetry of the QDM on the exciton transitions energies shown in
Fig.\,3(a) in dependence on (center-to-center) dot distance $d$.

At short distances $d$ the tunnelling exceeds both the Coulomb
energy and the asymmetry, and the exciton states are well
described in terms of one-particle states. Like in symmetric
QDMs~\cite{Szafran01}, optically active states
$|1\rangle=|SS\rangle$ and $|4\rangle=|AA\rangle$ are formed from,
respectively, symmetric and antisymmetric electron and hole
states, while the other two, $|2\rangle=|SA\rangle$ and
$|3\rangle=|AS\rangle$, remain dark. As $d$ increases, the QDM
asymmetry and the Coulomb interaction mix these symmetric
combinations as is clearly seen from Fig.\,3(b). Finally, in the
limit of large $d$ the two QDs become isolated (no tunnelling) and
bright states are $|1\rangle=|LL\rangle$ and
$|2\rangle=|RR\rangle$ formed from electrons and holes both
localized on the left and on the right dot, respectively. The
energy splitting between them is $\Delta_e+\Delta_h$, where
$\Delta_e$ ($\Delta_h$) is the electron (hole) asymmetric
splitting due to the slight difference between the QDs. The other
two, $|3\rangle$ and $|4\rangle$, are spatially indirect exciton
states which are dark and split off by the Coulomb energy $E_C$.
In contrast, in a symmetric QDM the two bright states would be
separated by the Coulomb energy~\cite{Szafran01}.

\begin{figure}[t]
\includegraphics*[width=8.0cm]{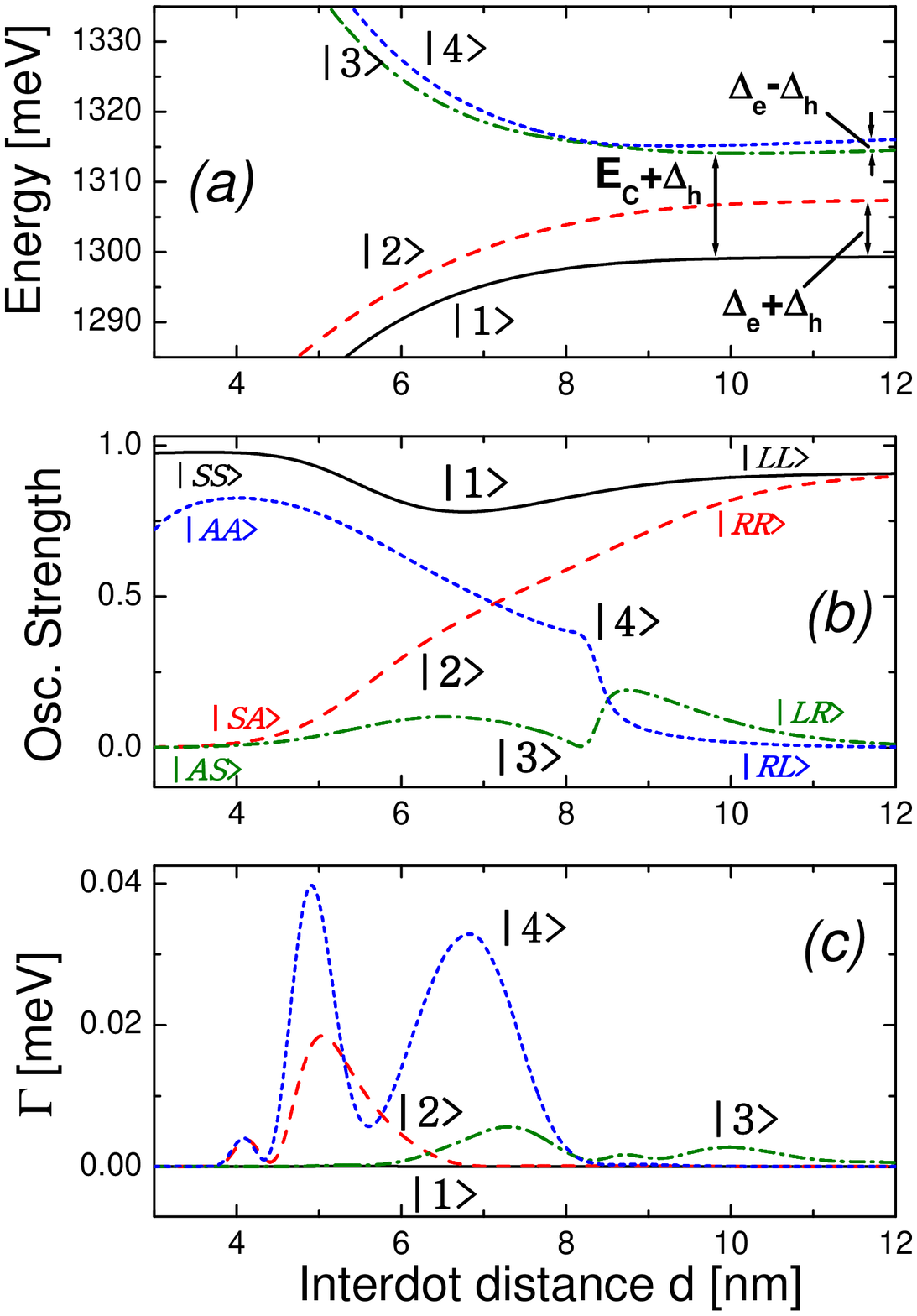}
\caption{ Exciton energies (a) and oscillator strength (b)
calculated within the four-level model of InGaAs QDM with account
for the Coulomb interaction and 2\% of asymmetry. (c): Exciton
dephasing rates of the QDM calculated at $T=10$\,K. The meaning of
$E_C$, $\Delta_e$ and $\Delta_h$ are given in the text.}
\end{figure}

The full linear polarization of a QDM is calculated up to second
order in the cumulant expansion, using the described four-level
excitonic model. It's Fourier transform, i.e., the absorption,
shown in Fig.\,4 contains four finite-width Lorentzian lines on
the top of broadbands. The width of the broadband is of the order
of the typical energy of phonons participating in the transitions,
$\omega_0\sim 2$\,meV. Thus, if two levels come close to each
other and the broadbands start to overlap, the ZPLs get
considerably wider, due to the real phonon-assisted transitions
between neighboring levels. In QDMs this important mechanism of
the dephasing is controlled by the tunnelling which induces a
level repulsion at short interdot distances.

\begin{figure}[t]
\includegraphics*[width=8cm]{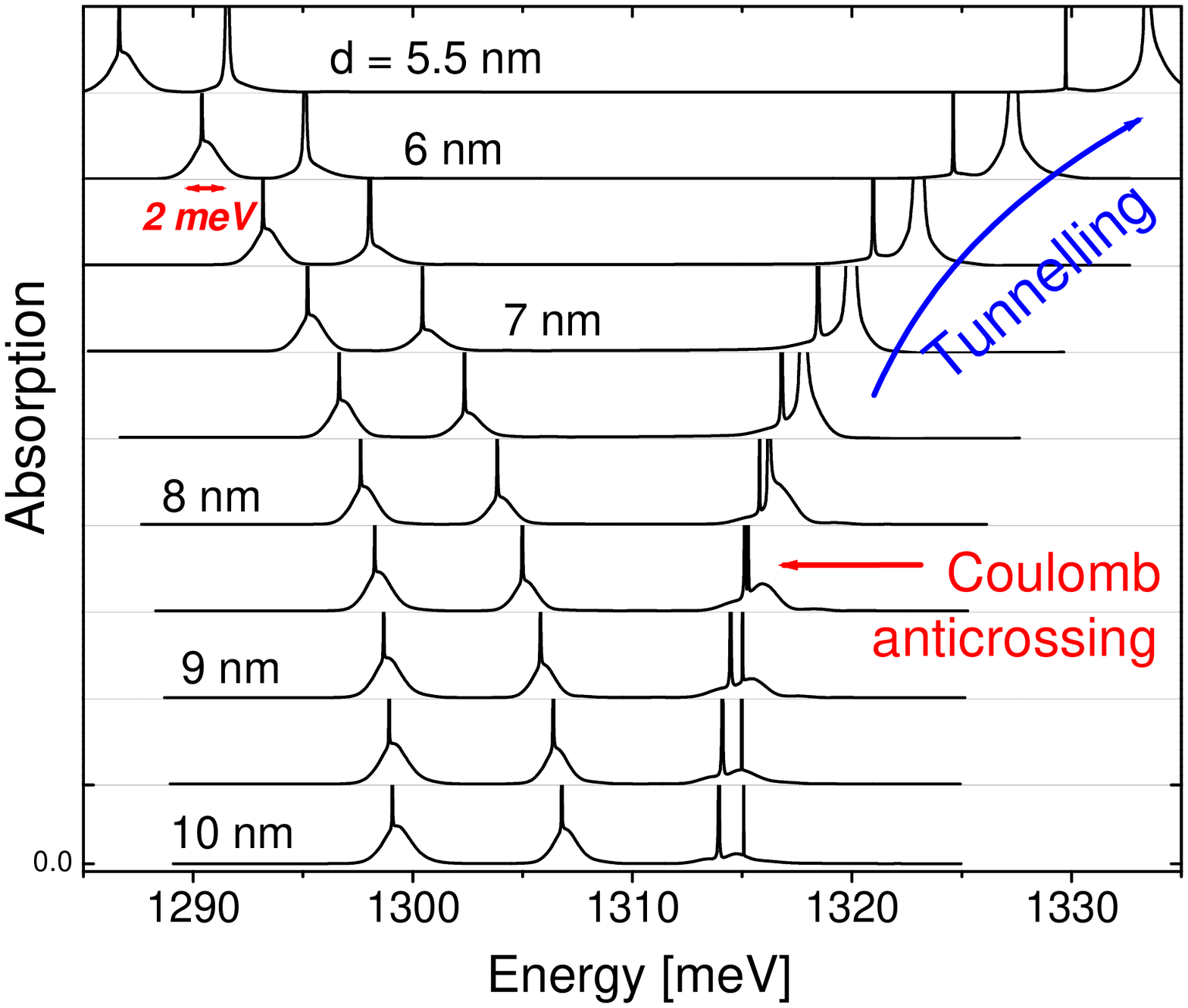}
\caption{ Absorption spectrum (linear scale) of an asymmetric
InGaAs QDM calculated at $T=10$\,K for different interdot
distances $d$. The peaks of the ZPLs are truncated.}
\end{figure}

However, there is another effect which leads to a quite unexpected
result: {\it Coulomb anticrossing}. As $d$ increases, the two
higher levels come close to each other and should exchange phonons
more efficiently. Nevertheless, when the anticrossing is reached
at around $d=8$\,nm (Fig.\,4), the ZPL width suddenly drops and
never restores at larger $d$. This is due to a change of the
symmetry of states, owing to the Coulomb interaction. At $d>8$\,nm
states $|3\rangle$ and $|4\rangle$ become more like $|LR\rangle$
and $|RL\rangle$, respectively [See Fig.\,3(b)], and the
exciton-phonon matrix element between them vanishes by symmetry.
Thus, real transitions between states $|3\rangle$ and $|4\rangle$
are not allowed any more and their dephasing is only due to
virtual transitions into states $|1\rangle$ and $|2\rangle$.

\begin{figure}[b]
\includegraphics*[width=7.5cm]{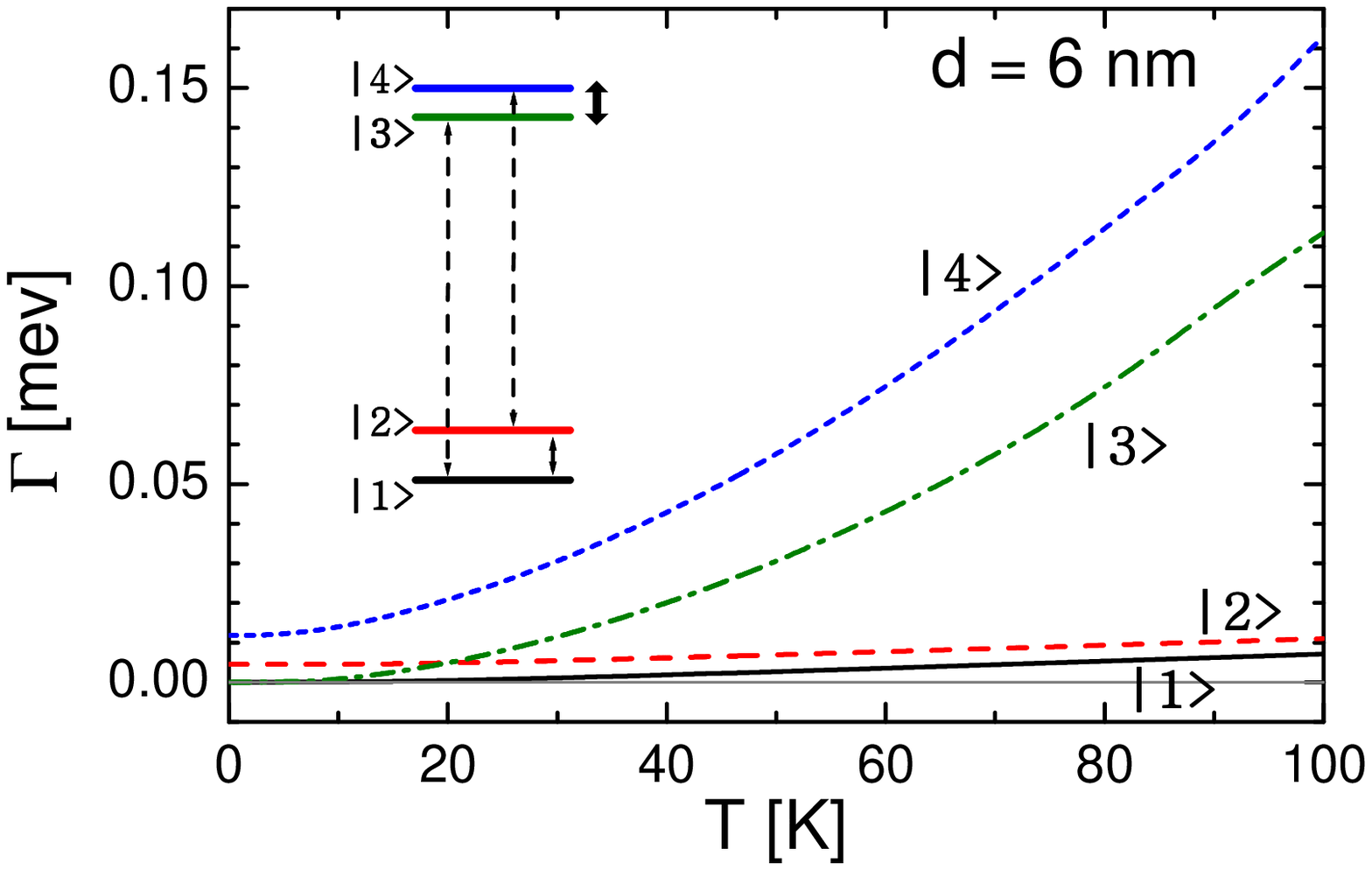}
\caption{ Temperature dependence of the ZPL widths $\Gamma_n$ for
the lowest four exciton levels in $d=6$\,nm QDM. Inset: real
(virtual) phonon-assisted transitions between exciton states shown
schematically by full (dashed) arrows. }
\end{figure}

Dephasing results for all four states are summarized in
Fig.\,3(c). Apart from the features already discussed, there are
also oscillations in $\Gamma$ clearly seen for levels $|2\rangle$
and $|4\rangle$ in the region between $d=4$\,nm and 8\,nm. For an
explanation, note that the matrix elements $M_{\bf q}^{nm}$,
Eq.(\ref{M}), are Fourier transforms of the electron (hole)
probabilities. When located in different QDs, they carry a factor
of $\exp(iqd)$. Since the typical phonon momentum participating in
real transitions is $q_0\sim 1/L_z$, one could expect that
$\Gamma$ has maxima spaced by a length of order $L_z$.

The temperature dependence of $\Gamma_n$ is shown in Fig.\,5. At
$d=6$\,nm, levels $|3\rangle$ and $|4\rangle$ are already close to
each other, and real phonon-assisted transitions between them are
possible. As a result, the dephasing rates grow quickly with
temperature. At the same time, levels $|1\rangle$ and $|2\rangle$
are far from each other and real transitions are suppressed.
Still, virtual transitions contribute everywhere, with no strong
dependence on level energies.

In conclusion, in the present microscopic approach to the
dephasing in quantum dot molecules, we go beyond Fermi's golden
rule and quadratic coupling model~\cite{Muljarov04}, taking into
account both real and virtual phonon-assisted transitions between
exciton levels on equal footing. While the dephasing in single
quantum dots is mainly due to virtual transitions, in quantum dot
molecules real transitions dominate the dephasing. We show that
the broadening of the zero-phonon lines calculated for a few
lowest exciton states depends strongly on interdot distance (via
tunnelling), electron-hole Coulomb interaction, and asymmetry of
the double-dot potentials.

Financial support by DFG Sonderforschungsbereich 296, Japan
Society for the Promotion of Science (L-03520), and the Russian
Foundation for Basic Research (03-02-16772) is gratefully
acknowledged.


\begin{thebibliography}{10}
\bibitem{note1} Both lengths are connected to each other through the electron
(hole) potential depth $U_{e(h)}$: $l_e^4m_e U_e=l_h^4m_h U_h$;
$U_e/U_h=7/3$~\cite{Landolt}.
\end{thebibliography}
\end{document}